\documentstyle[11pt,moriond,epsfig]{article}
\def\jour#1#2#3#4{{#1} {\bf #2}, #3 (#4)}
\def\modsubsection#1{\vskip -20pt\vphantom{p}\subsection{#1}
   \vphantom{p}\vskip -15pt}
\font\sevenrm=cmr7
\font\fiverm=cmr5

\def\ssr{\em Space Sci. Rev.}
\def\apj{\em Astrophys. J.}
\def\apjl{\em Astrophys. J. Lett.}

\def\apss{\em Astrophys. Sp. Sci.}

\def\phrep{\em Phys. Reports}
\def\aap{\em Astron. Astrophys.}
\def\annrev{\em Ann. Rev. Astron. Astrophys.}
\def\nat{\em Nature}
 
\def\pasp{\em Pub. Astron. Soc. Pacific}
\def\teq#1{$\, #1\,$}
\def\dover#1#2{\hbox{${{\displaystyle#1 \vphantom{(} }\over{
   \displaystyle #2 \vphantom{(} }}$}}

\def\eSN{E_{\hbox{\fiverm SN}}}
\def\Vsk{u_{\hbox{\sevenrm sk}}}
\def\Rsk{r_{\hbox{\sevenrm sk}}}
\def\tSNR{t_{\hbox{\fiverm SNR}}}

\def\Emax{E_{\hbox{\sevenrm max}}}
{\catcode`\@=11                                                  
\gdef\SchlangeUnter#1#2{\lower2pt\vbox{\baselineskip 0pt\lineskip0pt    
\ialign{$\m@th#1\hfil##\hfil$\crcr#2\crcr\sim\crcr}}}}           
\def\gtrsim{\mathrel{\mathpalette\SchlangeUnter>}}               
\def\lesssim{\mathrel{\mathpalette\SchlangeUnter<}}    

\def\figcaption#1{\vbox{\smallskip \baselineskip10.7pt 
   \footnotesize #1\smallskip}}
\begin{document}
\vspace*{4cm}
\input{psfig.tex}
\title{GAMMA-RAY PRODUCTION IN SUPERNOVA REMNANTS}

\author{Matthew G. BARING \footnote[1]{Compton Fellow, USRA.
   \hskip 15pt Email: \it Baring@lheavx.gsfc.nasa.gov}}

\address{Laboratory for High Energy Astrophysics, Code 661,\\
NASA/Goddard Space Flight Center, Greenbelt, MD 20771, U.S.A.}

\maketitle\abstracts{
Supernova remnants are widely believed to be a principal source of
galactic cosmic rays, produced by diffusive shock acceleration in the
environs of the remnant's expanding shock.  This review discusses
recent modelling of how such energetic particles can produce gamma-rays
via interactions with the remnants' ambient interstellar medium,
specifically via neutral pion decay, bremsstrahlung and inverse Compton
emission.  Predictions that relate to the handful of associations
between EGRET unidentified sources and known radio/optical/X-ray
emitting remnants are summarized.  The cessation of acceleration above
1 TeV - 10 TeV energies in young shell-type remnants is critical to
model consistency with Whipple's TeV upper limits; these observations
provide important diagnostics for theoretical models.
}

\section{Introduction}
\label{sec:intro}

For many years now it has been a common perception that supernova
remnants (SNRs) are a principal, if not the predominant, source of
galactic cosmic rays (e.g. see Lagage and Cesarsky\cite{lg83}) up to
energies of around \teq{\sim 10^{15}}eV, where the so-called {\it knee}
in the spectrum marks its deviation from almost pure power-law
behaviour (e.g. see Hillas\cite{hill84} for a depiction of the
spectrum).  Such cosmic rays are presumed to be produced by diffusive
shock (Fermi) acceleration in the environs of the shock that is
initiated by the impact of the supernova ejecta on the surrounding
interstellar medium.  The convenience of a SNR origin of cosmic rays
below the knee is founded in (i) the appropriateness of their ages
(between 100 and \teq{10^5} years) and sizes for permitting the
diffusive process to accelerate up to such high energies (see below and
the discussion in ref.~[1]), (ii) they have the necessary power to
amply satisfy cosmic ray energetics requirements, and (iii) that
current estimates of supernova rates in our galaxy can adequately
supply the observed cosmic ray density (e.g. see the review of
Blandford and Eichler\cite{be87}).

The evidence for cosmic ray acceleration in remnants is, of course,
circumstantial.  Nevertheless, the ubiquity of polarized, non-thermal
radio emission in remnants (e.g. see references in the SNR compendium
of Green\cite{dgreen95}) argues convincingly for efficient acceleration
of electrons if the synchrotron mechanism is assumed responsible for
the emission.  X-rays also abound in remnants, and are usually
attributed to thermal emission from shock-heated electrons (because of
the appearance of spectral lines, e.g. see Borkowski et
al.\cite{bsbs96} for Cas A; see also the review of Ellison et
al.\cite{ell94}).  The striking spatial coincidence of radio and X-ray
images of remnants (e.g Tycho\cite{be87} and SN1006; see Keohane et.
al.\cite{kra96} for a radio/X-ray correlation analysis for Cas A)
suggests that the same mechanism is responsible for emission in both
wavebands.  This contention has recently received a major boost with
the discovery of non-thermal X-ray emission in SN1006 by Koyama et
al.\cite{koy95}, which implies the presence of non-thermal electrons at
super TeV energies (see Reynolds\cite{rey96} for a detailed
description).  In addition, very recent ASCA spectra (Keohane et
al.\cite{keo96}) for the remnant IC 443 and RXTE observations of Cas A
(Allen et al.\cite{all97}) exhibit non-thermal X-ray contributions,
adding to the collection of TeV electron-accelerators.  A nice review
of radio and X-ray properties of SNRs is given in Ellison et
al.\cite{ell94}.

An offshoot of cosmic ray acceleration in SNRs is that such energetic
particles can produce gamma-rays via interactions with the ambient
interstellar medium (ISM).  Although no unequivocal evidence for
gamma-ray emission from isolated supernova remnants exists, Esposito et
al.\cite{esp96} presents a handful of associations between unidentified
EGRET sources (at above 100 MeV) in or near the galactic plane and
known radio/optical/X-ray-emitting (relatively young) remnants,
providing ample motivation for exploring the possibility of high energy
emission from SNRs.  Such associations are suggestive\cite{sd95}, but
suffer from the large uncertainty\cite{esp96} in position location of
EGRET sources, of the order of 0.5--1 degrees, i.e. the size of typical
nearby remnants (see the images depicted in Figure~1).  Hence a definitive
connection between {\it any} of these gamma-ray sources and the young
SNRs is not yet possible.  The situation is complicated by the presence
of a pulsar (PSR B1853+01) in the field\cite{esp96,djm97} of the 95\%
confidence contour of the EGRET source 2EG J1857+0118.  Such a pulsar
(or its wind nebula) could easily spawn the observed gamma-ray emission
(which has so far yielded no evidence of pulsation\cite{thom94}), and
the conceivability that pulsars could be responsible for most
unidentified EGRET sources near the galactic plane\cite{kc96}
(discussed in the review by Grenier in this volume) currently inhibits
any assertions stronger than just suggestions of a remnant/EGRET source
connection.

In this paper, the handful of SNR gamma-ray emission models that invoke
shock acceleration and have been developed in the last four years are
reviewed.  This field began in earnest (following limited early
work\cite{hl75,chev77}) with the seminal paper of Drury, Aharonian and
V\"olk\cite{dav94}, who computed the photon spectra expected from the
decay of neutral pions generated in collisions between
shock-accelerated ions and cold ions in the ISM.  Their work used a
two-fluid approach to shock acceleration.  Since, then there has been a
small flurry of activity, with different groups using alternative
approaches, and extending the considerations to include bremsstrahlung and
inverse Compton emission.  Following a brief summary of supernova
remnant expansion properties that are relevant to the acceleration and
ultimate energies of cosmic rays, the various models of gamma-ray
production will be discussed.  Then the focus will turn to very recent
work on the relevance of non-linear effects in shock acceleration
theory (discussed in the review paper of Baring\cite{bar97} in this
volume) to the SNR $\gamma$-ray spectra.  These effects describe the
dynamical influence of the accelerated cosmic rays on the shocked
plasma at the same time as the non-uniformities in the fluid flow force
the distribution of the cosmic rays to deviate from pure power-laws.
The recent TeV upper limits obtained\cite{less95} by the Whipple
air \v{C}erenkov detector (discussed by Weekes, this volume) play a
very significant role in SNR $\gamma$-ray models; contrary to common
perception, the various models can be comfortably fine-tuned to
accommodate this observational data.  Finally, recent work on broad-band
approaches to SN1006 and W44 is addressed, indicating the myriad of
possibilities for potential $\gamma$-ray-emitting supernova remnants.

%
\vskip+0.5truecm
\centerline{\hskip 1.0truecm\psfig{figure=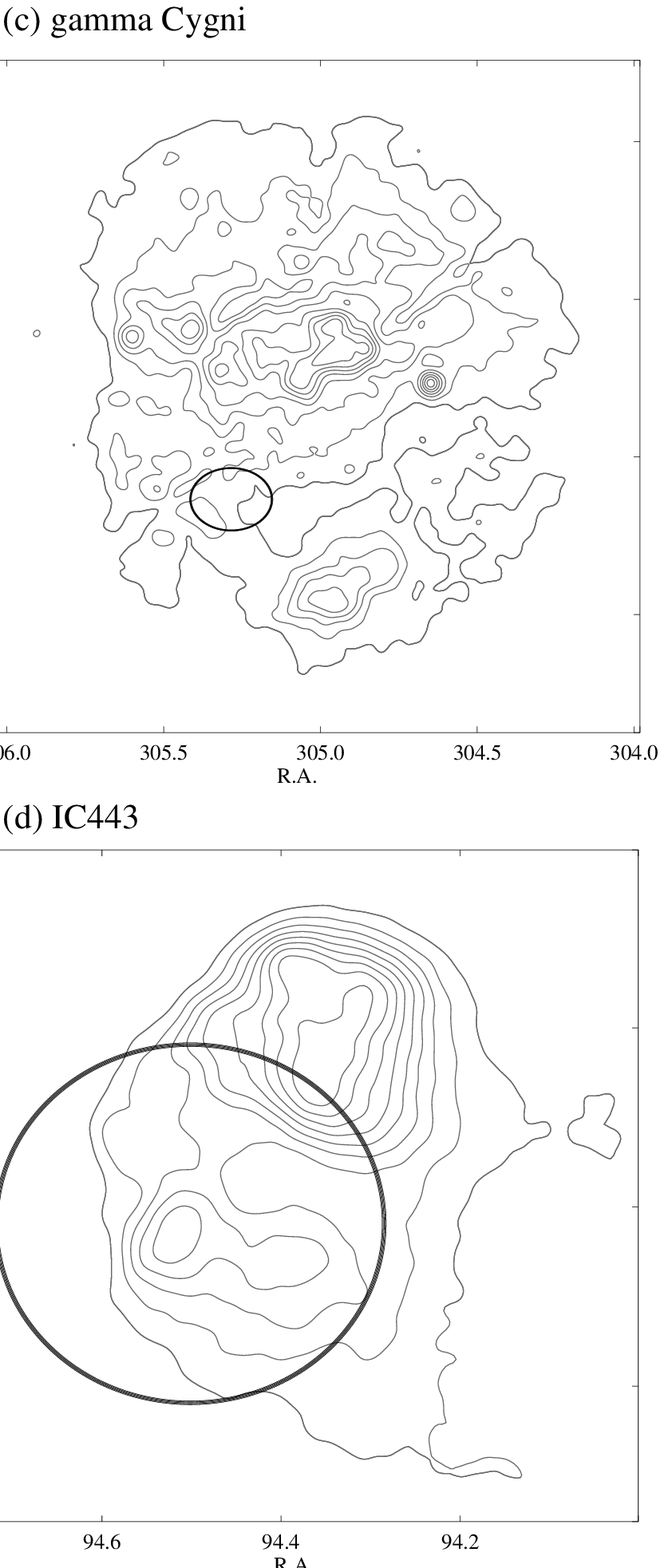,width=7.85truecm}
 \hskip -2.0truecm\psfig{figure=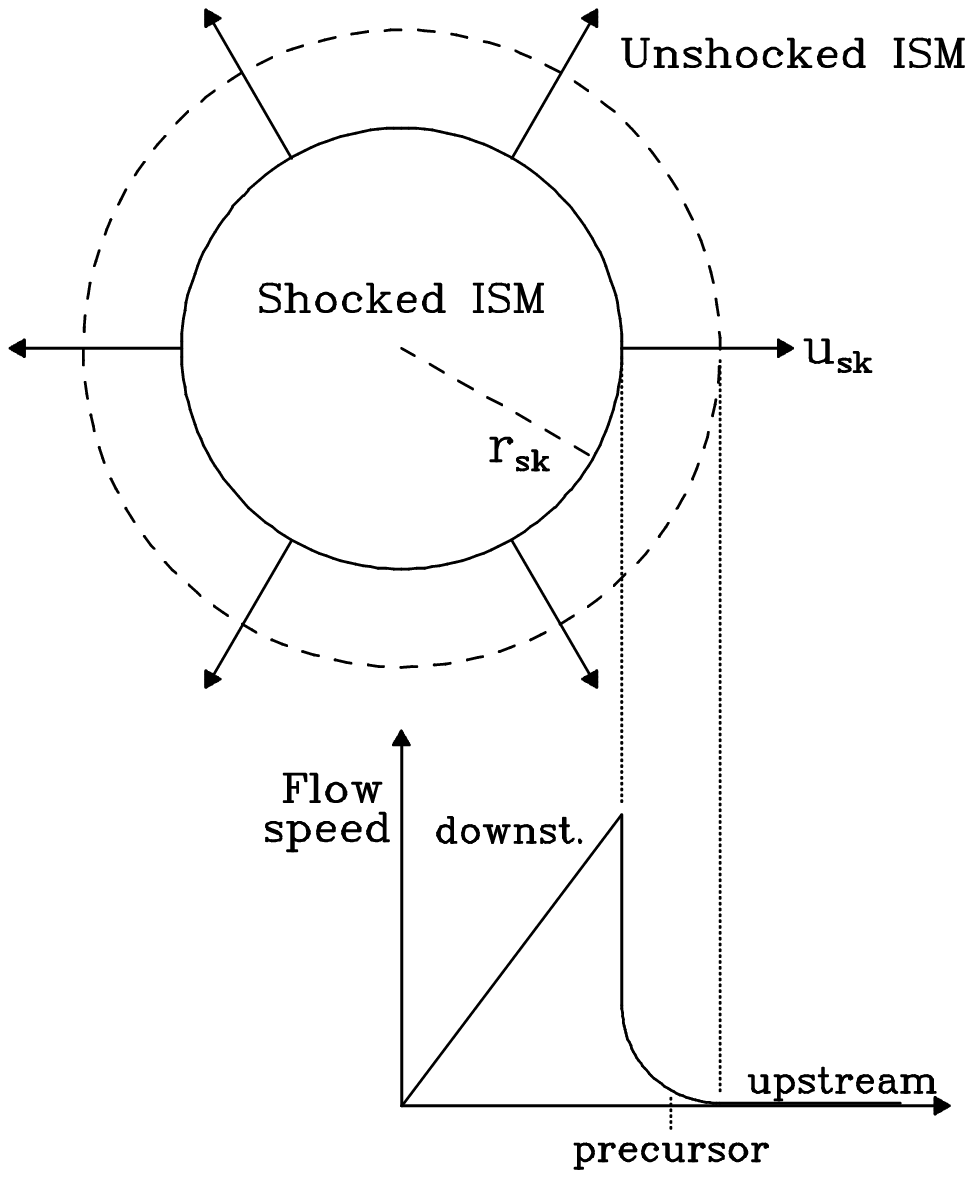,width=8.85truecm}}
\vskip-0.0truecm
\figcaption{Figure~1:  On the left are X-ray/gamma-ray images of the
supernova remnants $\gamma$ Cygni and IC443, as presented in Esposito
et al.\cite{esp96}.  These consist of X-ray contours from the ROSAT
telescope's HRI, and the 95\% confidence contours of emission above 100
MeV in associated EGRET unidentified sources.  On the right is a
schematic depiction of an expanding supernova remnant (in the Sedov
phase) together with its flow velocity spatial profile in the
observer's frame (downstream material with higher speeds is to the left
of the shock discontinuity).}

\section{Supernova Remnant Expansions}
\label{sec:expand}

It is instructive to review briefly the current conception of a
relatively young supernova remnant expansion.  This age, which is
broadly categorized as being between around 100 years and several
thousand years, is appropriate to $\gamma$-ray emission models for the
five EGRET source associations listed in Esposito et al.\cite{esp96}.
The low end of this range defines the termination of the epoch of
approximately free expansion and the gradual onset of the {\it Sedov}
(or scaling) phase of remnant evolution, during which the remnant
decelerates as it sweeps up the surrounding interstellar medium.
During its expansion, the remnant has a propagating forward shock
initiated by the slamming of the supernova ejecta into the ISM outside,
producing an adiabatically-expanding interior composed of shocked ISM
(see Figure~1 for a depiction).  It is during the Sedov phase that this
shock is most efficient at depositing its energy into cosmic rays with
subsequent conversion to $\gamma$-rays:  Drury et al.\cite{dav94} found
that the $\gamma$-ray emissivity indeed peaks during this part of the
SNR evolution.  At late stages, the Sedov phase has
weakened the shock considerably, and eventually it becomes radiative
after tens of thousands of years, i.e. in the post-gamma-ray epoch.

The standard Sedov solution (e.g. Lang\cite{lang80}) for an explosion
of initial energy \teq{\eSN}, in a gas of density \teq{\rho= 1.4 n_1
m_p} for number density \teq{n_1} (the
\teq{1.4} factor accounts for the cosmic abundance of helium and
heavier elements) is
\begin{equation}
\Vsk =\dover{2}{5} \xi \biggl( \dover{\eSN}{\rho} \biggr)^{1/5}
   \tSNR^{-3/5}\ ,\quad
\Rsk = \xi \biggr( \dover{\eSN}{\rho} \biggl)^{1/5} \tSNR^{2/5} \ ,
 \label{eq:Sedov}
\end{equation}
for the shock speed \teq{\Vsk} and radius \teq{\Rsk}, as a function of
the SNR age, \teq{\tSNR}.  Here \teq{\xi\approx 1.15}.  The onset of
the Sedov phase occurs smoothly over a range of times about some time
\teq{t_{\rm Sed}}, which is defined roughly by when the mass of the
supernova ejecta becomes comparable to that of the ISM it displaces
i.e. when \teq{\Vsk} equals the free-expansion speed
\teq{\sqrt{2\eSN}/M_{\rm ej}}.  This yields \teq{t_{\rm Sed}\sim 0.1
\eSN^{-1/2}\, M_{\rm ej}^{5/6}\, (n_1m_p)^{-1/3}}, and therefore values
of 50 to 300 years for \teq{\eSN\sim 10^{51}}ergs.  During the Sedov
phase, the shock speed is typically 200--3,000 km/sec.

A property of Fermi acceleration that is crucial to the interpretation
of spectra via theoretical models is the maximum energy of cosmic rays
permitted in a remnant at its present age.  Obviously, since the
first-order Fermi process involves monotonic evolution of energies with
time, the acceleration time to a given energy cannot exceed the age of
the remnant.  Thereby, one bound to the maximum energy can be
obtained.  From standard particle diffusion theory, the oft-quoted form
for the acceleration time \teq{\tau_a}, to a given energy \teq{\Emax}
was derived by Forman and Morfill\cite{fm79}.  Their formula is applied
here to plane-parallel shocks, but can be easily adapted to shocks of
arbitrary obliquity\cite{jok87,ebj95}.  For \teq{r=u_1/u_2}, their
result gives an acceleration time
\begin{equation}
  \tau_a\; \sim\; 10^4\; \dover{r(1+r)}{r-1}\, \dover{\eta}{Q}
  \biggl( \dover{B_1}{3 \mu {\rm G}} \biggr)^{-1}
  \biggl( \dover{\Emax}{10^{2} {\rm TeV}} \biggr)
  \biggl( \dover{\Vsk}{10^3 {\rm {km~s^{-1}} } } \biggr)^{-2} {\rm years}.
 \label{eq:taua}
\end{equation}
Equating this to \teq{\tSNR} gives an upper limit to the maximum energy
of acceleration \teq{\Emax}.  Acceleration is also limited in the
spatial domain: the typical diffusion scale \teq{d_{\rm
diff}\sim\kappa_{\rm max}/\Vsk = \eta r_{\rm g,max} c/(3 \Vsk )=\eta\,
\Emax /(3 eQ\, \Vsk\, B_1)} must be somewhat less than the current
remnant size \teq{\Rsk}, due to escape upstream of the shock in a spherical
geometry.  Here \teq{\eta =\lambda /r_g} for mean free paths
\teq{\lambda}.  The relevant size of the acceleration region can
further be constrained by the presence of dense neutral regions or
incompletely ionized media\cite{ddk96} in the environs of the shock,
since such regions strongly suppress wave generation, and therefore
also the Fermi process.  For young SNRs early in the Sedov epoch,
\teq{d_{\rm diff}\lesssim 0.1 \Rsk}, a result that is in accord with
the non-linear analysis of Drury et al.\cite{dav94} (see below),
and the derived spatial/temporal limit on \teq{\Emax} is
\begin{equation}
  \Emax\; \sim\; 0.1 \dover{Q}{\eta}
  \biggl( \dover{B_1}{3\mu {\rm G}} \biggr)
  \biggl( \dover{\Vsk}{10^3 {\rm {km~s^{-1}} } } \biggr)^2
  \biggl( \dover{\tSNR}{{10^3 {\rm yr}} } \biggr)\ \ {\rm TeV}\; .
 \label{eq:emax}
\end{equation}
This generally yields \teq{\Emax} (which peaks in the Sedov phase) for
ions that are in the 1--30 TeV range per nucleon, depending on the
remnant parameters.  Note that effects of shock obliquity can act to
increase the maximum energy\cite{jok87}, which becomes crucial to
easing constraints\cite{rey96} imposed by the observations of
non-thermal X-rays in SN1006.  As discussed in Baring\cite{bar97},
effecting such increases in \teq{\Emax} compromises the efficiency of
generating populations at these high energies.

\section{Gamma-Ray Production Models}
\label{sec:gamma}

In the generation of gamma-rays in SNR environments, there are a
handful of processes that are relevant, spawned by the collisions of
shock-accelerated electrons and ions with the cold ISM.  Foremost among
these is the decay of neutral pions, which are formed in hadronic
collisions \teq{pp\to p\pi^0X} etc., into two gamma-rays.  Due to the
isotropy of decay in the pion rest frame, the kinematics of this decay
yields\cite{steck71} a photon spectrum that is symmetric about
\teq{m_{\pi}/2\approx 67}MeV, an unmistakeable signature of the
production of pions in astrophysical systems.  Supernova remnant models
of pion production and decay use some variant of a hybrid approach
(e.g. see Dermer\cite{derm86}), where low energy pion creation (for
shock-accelerated proton momenta \teq{p_p} below around 3 GeV/c) is
mediated by an isobaric state \teq{\Delta (1232)}
(Stecker\cite{steck71}), or a collection of different states, and the
complexities of pion creation at high energies (for \teq{p_p\gtrsim 10}
GeV/c) is described by some adaptation\cite{sb81,tn83} of Feynman
scaling.  Among the non-hadronic processes that are pertinent to SNR
gamma-ray models is inverse Compton scattering by relativistic
electrons\cite{rl79} of soft (i.e. low energy) photon fields such as
the cosmic microwave background or infrared (IR) backgrounds local to
the remnants.  These contribute both X-rays and gamma-rays to the
emission.  At the same time, it is possible that electron synchrotron
radiation, which can extend from radio up to X-ray energies, can act as
seeds for the inverse Compton process.  In addition, bremsstrahlung
between relativistic \teq{e} and ions and ISM electrons can generate
gamma-rays, and form a potentially important part of gamma-ray models.
Finally, Coulomb collisions, which are insignificant when the remnant
environment is collisionless (for most of its lifetime), can help to
redistribute electron and proton energies below 100 MeV when the
remnant age exceeds \teq{10^4}years\cite{ssd97}.  These processes are
addressed in one or other of the various papers discussed below.

\modsubsection{The Two-Fluid Model of Drury et al.}

The first gamma-ray emission model for supernova remnants presented in
the recent wave of interest was the seminal work of Drury, Aharonian
and V\"olk\cite{dav94}, who computed the photon spectra and fluxes
expected from the decay of neutral pions generated in collisions
between shock-accelerated ions and cold ions in the ISM; they neglected
the other electromagnetic processes mentioned just above.  Their work
used the {\it two-fluid} approach\cite{dv81,drury83} to shock
acceleration (discussed briefly in Baring\cite{bar97}), treating the
cosmic rays and thermal ions as separate entities (electrons go along
for the ride).  This technique explores the hydrodynamics of shocked
flows taking into account these two components, obtaining solutions
that conserve particle number, momentum and energy fluxes (per solid
angle in their spherically symmetric application to SNRs); it therefore
describes, in a fashion, non-linear effects\cite{bar97} in shock
acceleration.  The two-fluid approach is extremely useful for
time-dependent applications, and therefore is appropriate for SNRs, but
generally contains little or no self-consistent spectral information
(but see the recent work of Malkov and V\"olk\cite{mv96}), in contrast
to the Monte Carlo simulations discussed in Baring\cite{bar97}.   The
model of Drury et al.\cite{dav94} built on an earlier two-fluid
analysis\cite{dmv89} of remnant evolution, which assumed particle
diffusion in the Bohm limit\cite{bar97} where diffusion mean free paths
\teq{\lambda} are comparable to the ion gyroradii \teq{r_g}.  They were
able to map the gamma-ray luminosity and shock profile evolution, and
determined that the luminosity peaked in the Sedov phase and was more
or less constant throughout it.  This is in accord with a maximal shock
dissipation when the supernova ejecta is being compressed and
significantly decelerated by the ISM.

Drury et al.\cite{dav94} noted that their model required a high target
density (\teq{>100}cm$^{-3}$) to match the EGRET flux, a situation that
seems inevitable if pion decay emission dominates gamma-ray spectral
formation.  They observed, furthermore, that the remnants should become
limb-brightened with age, an effect that arises because the shock
weakens with time so that the dominant gamma-ray flux is always
``tied'' somewhat to a region near the shock that is sampled by lower
particle energies.  This attachment naturally requires the ionic mean
free path \teq{\lambda} to be an increasing function of the momentum,
so that particles of lower energies remain closer to the shock.  While
such a limb-brightening is seen in radio and X-ray images of remnants
(e.g. Tycho and SN1006), higher angular resolution capability will be
needed in gamma-ray telescopes before its existence, or otherwise, can
be probed at high energies.  Drury et al. noted that the flow profile
modifications (see Figure~1) lead to a precursor width of about 1/10th
of the shock radius.  Despite its strengths, their work has two key
spectral deficiencies.  The first is that they are forced to {\it
assume} a canonical test-particle power-law form for the ion
populations, and hence do not describe self-consistently the intimate
relationship (discussed below, and by Baring\cite{bar97}) between
non-linear modification of the flow profile and spectral curvature.
Secondly, they assumed that the particles would be accelerated to at
least 100 TeV and did not incorporate physical limits such as in
Eq.~(\ref{eq:emax}) to the acceleration mechanism.  This latter
omission, promoted observational investigations by the Whipple
collaboration that produced upper limits in the TeV energy
range\cite{less95} that contradicted the Drury et al. predictions.
While this conflict has been proposed as a failure for shock
acceleration models of SNRs, realistic choices of \teq{\Emax} in
Eq.~(\ref{eq:emax}) actually yield model spectra that are quite
compatible with the observational constraints.

\modsubsection{More Recent Models}

The next substantial development of models of gamma-ray emission from
SNRs was the work of Gaisser, Protheroe and Stanev\cite{gps97}.  Their
computation of emission fluxes and luminosities for bremsstrahlung,
inverse Compton scattering, and decay of \teq{\pi^0}s produced in
hadronic collisions expanded beyond the consideration of Drury et al.
However, they omitted consideration of non-linear shock dynamics in any
form, did not treat time-dependence, and assumed just test-particle
power-law distributions of protons and electrons, with arbitrary
relative abundances (i.e. $e$/$p$ ratios) in cosmic-ray populations.
In their model, the inverse Compton scattering used both the microwave
background and an infrared/optical background field local to the SNRs
as seed soft photons.  Their bremsstrahlung component was due to cosmic
ray electrons colliding with ISM protons.  The Gaisser et al.  particle
populations extended as power-laws beyond 30 TeV, with cutoffs only due
to inverse Compton (IC) cooling, so that they included no treatment of
spatial or temporal limits to acceleration.  Consequently, they have
exactly the same problem as Drury et al.\cite{dav94} in that their
model violates the TeV upper limits obtained by Whipple.  At the same
time the large hard X-ray synchrotron fluxes that would result from
their model would violently conflict with observational limits.
Gaisser et al. observed that, since the inverse Compton spectrum is
intrinsically flatter than bremsstrahlung and pion decay spectra, this
component needs to be suppressed to accommodate the EGRET spectral
indices of the sources associated with $\gamma$ Cygni and IC443.  Hence
they imposed a high matter density (\teq{> 300}cm$^{-3}$) to enhance
bremsstrahlung and \teq{\pi^0} decay to IC flux ratio.  Such a goal can
also be achieved, without resorting to extremely dense ambient media,
by reducing the primary accelerated electron population.  Note that
\teq{\pi^{\pm}\to e^{\pm}} secondaries are always unimportant for the
SNR problem since the ion cooling time in pion production is much
longer than typical remnant ages.

Very recently Sturner, Skibo and Dermer\cite{ssd97} have developed a
time-dependent model, where they use the Sedov solution in
Eq.~(\ref{eq:Sedov}) for the expansion, and numerically solve
time-dependent equations for electron and proton distributions subject
to cooling by inverse Compton scattering, bremsstrahlung, \teq{\pi^0}
decay and synchrotron radiation.  Essentially they have included all
the radiation processes of Gaisser et al.\cite{gps97}, and have added
synchrotron emission to supply a radio flux.  One variant on all
previous work is their inclusion of Coulomb collisions, which they find
can contribute to the cooling of non-thermal ions and electrons below
100 MeV for remnants older than \teq{10^4} years.  Like Gaisser et al.,
the work of Sturner et al. is not really a shock acceleration model,
but rather a time-dependent particle evolution and
radiation emission approach.  They assume canonical power-laws like
Drury et al.\cite{dav94} and Gaisser et al.\cite{gps97}, and therefore
have no handle on how non-linear shock modification effects determine
the spectral index and curvature.  They do not include any non-linear
hydrodynamic effects as in Drury et al., and so omit any consideration
of dynamic modifications to the Sedov solution.  Sturner et al. obtain
gamma-ray emission that persists, as in Drury et al., into the
radiative phase of remnant evolution.  One feature of their model is
the dominance of inverse Compton emission, which intrinsically has a
flatter spectrum than either bremsstrahlung or pion decay radiation.
This arises because they generally opt to have the same energy density
in non-thermal electrons and protons, and therefore the
shock-accelerated electrons are much more populous than their proton
counterparts.  The inappropriateness of this from the point of view of
shock acceleration theory is discussed below.  Sturner et al.'s work
possesses a significant advance over the previous work by introducing
cutoffs in the distributions of the accelerated particles (actually
first done by De Jager and Mastichiadis\cite{djm97,mdj96}), which are
defined by the limits on the achievable energies in Fermi acceleration
discussed above.  Hence, given suitable model parameters, Sturner et
al. can accommodate the constraints imposed by Whipple's upper
limits\cite{less95} to $\gamma$ Cygni and IC 443.

\section{Spectral Effects of Non-Linear Shock Acceleration}
\label{sec:nonlinear}

The models addressed so far have many assets, but also two noteworthy
spectral limitations from the perspective of shock acceleration theory:
(i) they do not pin down the steepness of, nor describe curvature in, the
distribution of shock-accelerated ions and electrons, and therefore do
not self-consistently determine the particle populations, and (ii) they
possess no coherent prescription for the \teq{e}/\teq{p} ratio, an
abundance ratio that is crucial to the formation of the gamma-ray
spectrum.  These deficiencies arise because their considerations of
accelerated particle distributions are entirely test-particle in
nature; Drury et al. used the two-fluid approach to analyze
non-linear aspects of only the flow dynamics.  A treatment of these
non-linearities, which arise in supernova remnants because their shocks
are strong so that the generated cosmic rays are endowed with a
significant fraction of the total particle pressure, is essential for
more accurate modelling of emission spectra.  These spectral
limitations are naturally remedied by the Monte Carlo simulational
approach, as described in the accompanying review of
Baring\cite{bar97}.  This kinematic technique can self-consistently
model the feedback of the accelerated particles on the spatial profile
of the flow velocity, which in turn determines the shape of the
particle distribution.  To re-cap the discussion in Baring\cite{bar97},
the accelerated population pushes on the upstream plasma and
decelerates it before the discontinuity is reached, so that an upstream
{\it precursor} forms, in which the flow speed is monotonically
decreasing.  At the same time, the cosmic rays press on the downstream
gas, slowing it down too.  The overall effect is one where the total
compression ratio \teq{r}, from far upstream to far downstream of the
discontinuity, actually {\bf exceeds that of the test-particle
scenario}.  This situation, which results from the need of the flow to
increase \teq{r} to adjust for energy and momentum
escape\cite{eich84,ee84}, is illustrated in the left hand panel of
Figure~2, where the flow profile is depicted in the rest frame of the
shock: notice the inversion from the observer's perspective of the
profile that is depicted in Figure~1.

%
\vskip+0.5truecm
\centerline{\psfig{figure=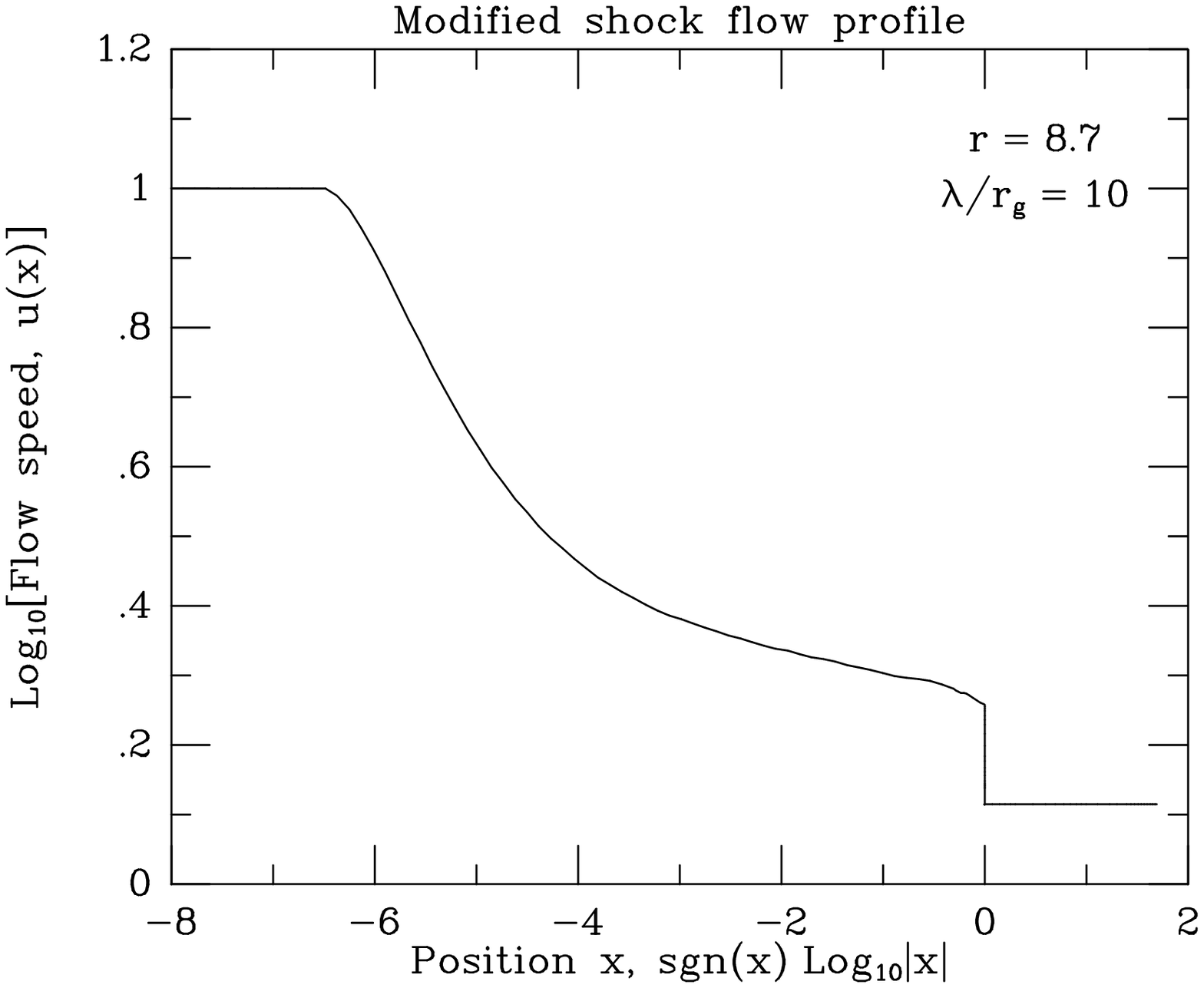,width=7.85truecm}
   \hskip 0.2truecm\psfig{figure=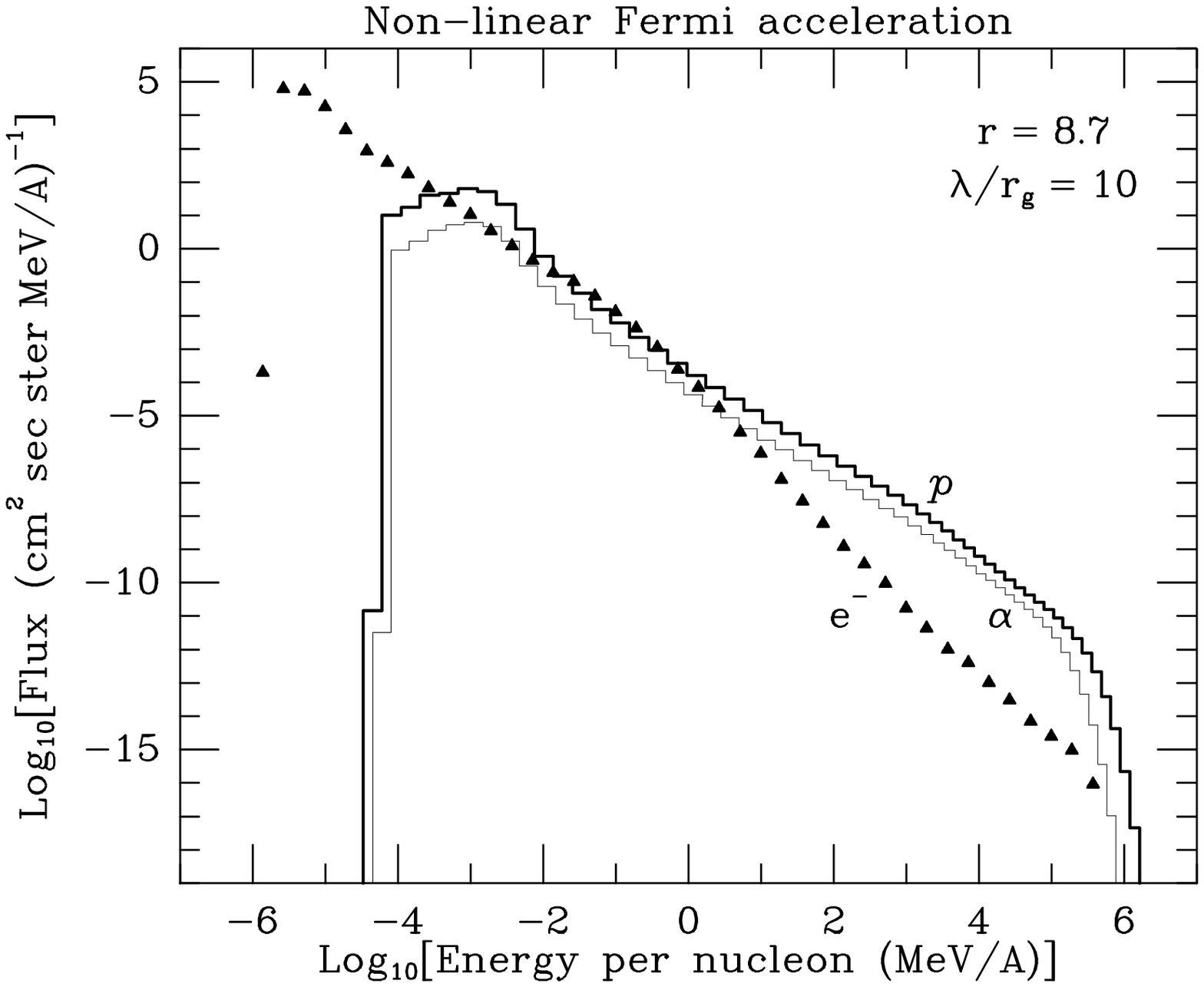,width=7.85truecm}}
\vskip-0.0truecm
\figcaption{Figure~2:  The velocity profile of fluid flow for a Monte
Carlo simulation of a non-linear shock that is modified by the cosmic
ray pressure is depicted on the left.  The abscissa represents the
position \teq{x} relative to the shock (\teq{>0} downstream) in units
of the mean free path of particles with speed \teq{u_1}.  Mostly the
abscissa is defined by the logarithmic form
\teq{\hbox{sgn}x\log_{10}\vert x\vert}, but it is linear in \teq{x} for
a small range near the shock. On the right are the particle
distributions (protons: heavy histogram, \teq{He^{2+}}:  light
histogram, electrons:  triangles) obtained by the simulation run that
generated the profile on the left.  In this instance, a
compression ratio of \teq{r=8.7} resulted.}

Typical distributions of particles that are accelerated in {\it
modified shock} flow profiles like that of Figure~2 are presented in
numerous papers\cite{eich84,ee84,je91,ebj96}: these resemble the right
hand panel of Figure~2, which is the distribution that was obtained
self-consistently\cite{bergg97} by the Monte Carlo simulation technique
in conjunction with the profile in the left hand panel.  The maximum
energies in these distributions were determined via the spatial
constraint that leads to Eq.~(\ref{eq:emax}).  Figure~2 therefore
embodies the intimate relationship between non-linear modification of
the flow profile and {\it upward} spectral curvature that is the
trademark\cite{eich84,ee84} of non-linear acceleration.  Such curvature
is a consequence of the use of a momentum-dependent mean free path
\teq{\lambda} in the Monte Carlo model.  When \teq{\lambda} is an
increasing function of momentum, an assumption that is supported by
inferences of particle diffusion from the Earth's bow shock and also in
hybrid plasma shock simulations\cite{gbse93}, the higher energy
particles sample large scales in their diffusion in the shock
environs.  Hence they experience a greater effective compression ratio
(see Figure~2), and consequently yield a flatter spectral index than at
low energies.  This curvature is important for gamma-ray emission
models, since it introduces enhancements in the TeV range by factors of
2--3 relative to the EGRET range; such increases can be the difference
between detection and non-detection by air \v{C}erenkov experiments
like Whipple, MILAGRO and CAT.

Gamma-ray emission spectra that are generated by the self-consistent
Monte Carlo approach to shock acceleration are depicted in Baring,
Ellison and Grenier\cite{beg97} and Baring et al.\cite{bergg97}.  The
former work focuses on just neutral pion decay emission, while the
latter also includes bremsstrahlung, synchrotron and inverse Compton
emission components for SNRs.  In both papers, the cessation of
acceleration above critical energies in the 1 TeV - 10 TeV range caused
by the spatial and temporal limitations of the expanding SNR shell
[i.e. according to Eq.~(\ref{eq:emax})] yields gamma-ray spectral
cutoffs, so that the resulting emission spectra appear to be consistent
with Whipple's TeV upper limits\cite{less95} to those EGRET's
unidentified sources that have SNR associations.  Hence, as in Sturner
et al.\cite{ssd97}, the Whipple upper limits pose no serious problem,
but rather now provide powerful diagnostic constraints to our
theoretical understanding.  For example, combining
Eqs.~(\ref{eq:Sedov}) and~(\ref{eq:emax}), the value of \teq{\Emax} is
quite insensitive (in the Sedov phase) to the SNR age so that
sufficiently low values of \teq{\Emax} can only be obtained for low
field strengths \teq{B_1} or supernova energies \teq{\eSN}.  This
requirement can be quite stringent if non-thermal X-ray emission also
constrains the field strength, as will be discussed below.  Note that
\teq{\Emax} always peaks early in the Sedov phase.  The cosmic ray
spectral curvature plays a vital role in determining the emission
fluxes as the spectrum rolls over just below the cutoff, an important
consideration for possible detections by future instrumentation such as
GLAST, VERITAS and Celeste.  Note that the Monte Carlo approach of
Baring et al.\cite{bergg97,beg97} is not time-dependent, a limitation
that may not be serious at all, given that the upstream precursor
scalelength of \teq{\sim 0.1\Rsk} that was obtained by Drury et
al.\cite{dav94} was used as simulational input for the scale on which
particles escape the shock.  On this scale, the effects of shock
curvature can be neglected.

Another prominent feature of the distributions in Figure~2 is the low
value of the electron to proton ratio above 1 GeV.  This strongly
contrasts the situation of Sturner et al.\cite{ssd97}, who have an
electron-dominated cosmic ray component.  The reason the Monte Carlo
simulations produce a minority of non-thermal electrons (even though
the thermal populations satisfy charge neutrality) is that both proton
and electrons are injected directly from thermal energies.  The
electrons, however, suffer from a suppression of injection due to a
potential absence of resonant waves\cite{bar97} to effect diffusion
until they reach around 1 MeV.  This property of plasmas is modelled by
Baring et al.\cite{bergg97} via an electron mean free path that exceeds
that of protons at energies below around 1 MeV.  For non-relativistic
particle speeds, electrons have shorter \teq{\lambda} than do protons
of comparable kinetic energies, and hence sample smaller compressions
in the modified shock profile.  The net result is that the electron
distribution is steep enough at low energies so as to render the
$e$/$p$ ratio much less than unity above 1 GeV.  This determination is
entirely consistent with the observation that electrons supply around
2\% of the cosmic ray population by number (e.g.  M\"uller et
al.\cite{mull95}), and also blends with limits on the local $e$/$p$
abundance ratio imposed when modelling the galactic gamma-ray
background\cite{hun97}.  The values of $e$/$p$ much greater than unity
in the Sturner et al.\cite{ssd97} model grossly violate these
observations and the understanding of wave properties in plasmas just
mentioned.  Future measurements of the unidentified sources by more
sensitive experiments in the 1--100 MeV range should constrain the
$e$/$p$ ratio.

\section{A Myriad of Possibilities}
\label{sec:bband}

While the focus here has been on gamma-rays from remnants, much can be
learned from studying other wavebands also.  This has been the approach
of Mastichiadis and De Jager\cite{mdj96,djm97}, who have examined the
remnants SN1006 and W44.  For SN1006, which has not been seen in
gamma-rays, they used\cite{mdj96} the recent observations\cite{koy95}
of non-thermal X-rays by ASCA to constrain the energy of electrons and
the magnetic field, interpreting the X-ray flux as being of synchrotron
origin.  This contention (see also Reynolds\cite{rey96}) assumes that
the steep X-ray spectrum is part of a rollover in the electron
distribution at energies around 100 TeV.  Using microwave and infrared
backgrounds appropriate to SN1006, Mastichiadis and De
Jager\cite{mdj96} predicted the resulting inverse Compton component in
gamma-rays, and determined that it would always satisfy the EGRET upper
bounds.  However, they concluded that TeV upper limits from experiments
like Whipple could potentially constrain the parameter \teq{\eta
=\lambda/r_g} in Eqs.~(\ref{eq:taua}) and~(\ref{eq:emax}) to values
signifying departure from Bohm diffusion (i.e. \teq{\eta\gg 1}),
otherwise the TeV flux would exceed that of the Crab nebula.  Pinning
the X-ray synchrotron spectrum determines \teq{\Emax^2 B} and also a
combination of \teq{B} and the electron density.  From
Eq.~(\ref{eq:emax}), \teq{\eta} therefore couples to \teq{B} and the
gamma-ray inverse Compton flux must anti-correlate with both \teq{B}
and \teq{\eta =\lambda /r_g}.

At the same time, De Jager and Mastichiadis\cite{djm97} have proposed
that the radio spectrum in W44 is too flat to be explained by a
shock-accelerated electron population.  This is not necessarily the
case, given the flat distributions (i.e. flatter than \teq{E^{-2}})
that can be attained at relativistic energies (see Figure~2), however
they have conjectured that the pulsar that is present in W44 may inject
electrons with the required distribution via its relativistic wind,
thereby circumventing the need to invoke Fermi acceleration at the
remnant's shock.  This opens up the possibility that plerionic sources
are distinct in their acceleration properties from non-plerionic ones,
despite displaying similar emission properties: W44 is a good candidate
for the unidentified EGRET source 2EG J1857+0118.  Add to this the fact
that the Vela pulsar has a wind nebula but no observable gamma-ray
emission associated with the outer extremity (i.e. shell) of its
remnant, and the picture is confused further.  De Jager and
Mastichiadis\cite{djm97} suggest that inverse Compton emission and
bremsstrahlung dominate the gamma-rays in W44, and invoke a high
$e$/$p$ ratio; the EGRET data and Whipple upper limits can still
accommodate a pion decay origin for gamma-rays and low $e$/$p$ ratios
without compromising the radio continuum.

These papers serves to underline the diversity of the handful of
remnants that are associated with EGRET unidentified sources.  Such a
diversity is also reflected in their morphological properties, their
optical/IR spectra and line emission, environmental densities, etc.,
and also the role of pulsars and plerionic contributions.  In short,
these remnants must be considered on a case-by-case basis.  Yet all
have proximity to dense molecular clouds of various sorts, providing a
strong clue to the reason for gamma-ray emission: a consensus that the
presence of dense molecular clouds may be the driving force and
signature of gamma-ray bright remnants seems to be emerging.  By the
same token, the gamma-ray emitters must be a minority of remnants
(perhaps mostly young) given that they cannot produce ions above around
a few TeV in profusion: remnants that provide cosmic rays up to the
knee must be a gamma-ray quiet majority.  Much remains to be explored
in this field, in particular the relationship between the clouds and
the shock parameters, the degree of ionization of the environment, the
precise location of the gamma-ray emission, differentiation between
plerion-driven and shock-powered gamma-ray sources, and the maximum
energies and relative abundances of the produced cosmic rays.  The
Whipple upper limits have not destroyed the hypothesis that SNR shocks
energize the particles responsible for the gamma-ray emission, but
rather are a blessing in disguise, providing a powerful tool for
constraining models.  The next generation of gamma-ray telescopes, with
better angular resolution and cumulatively-broad spectral range will
have a significant impact on this field, particularly in coordination
with X-ray and radio observations.

\vskip 9pt\noindent {\bf Acknowledgments:}
I thank my collaborators Steve Reynolds, Isabelle Grenier and Don Ellison, 
and also Okkie De Jager for many informative discussions about
supernova remnants and shock acceleration theory.  I also thank
Joe Esposito for providing the images used in Figure~1.


\end{document}